\documentclass[runningheads]{llncs}
\usepackage{arydshln}
\usepackage[T1]{fontenc}
\usepackage{graphicx}
\usepackage{xcolor}
\usepackage{hyperref}
\usepackage{color}

\usepackage[firstpage]{draftwatermark}

\SetWatermarkAngle{0}
\SetWatermarkText{\raisebox{12.5cm}{
\hspace{0.1cm}
\href{https://doi.org/10.5281/zenodo.15468864}{\includegraphics{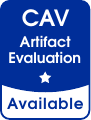}}
\hspace{10cm}
\includegraphics{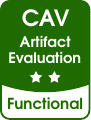}
}}

\begin{document}
\title{The rIC3 Hardware Model Checker}

\author{
    Yuheng Su\inst{1,2} \and
    Qiusong Yang\inst{1,3}\thanks{Qiusong Yang is the corresponding author.} \and
    Yiwei Ci\inst{1} \and
    Tianjun Bu\inst{1,2}\and
    Ziyu Huang\inst{4}}
\authorrunning{Y. Su et al.}
\institute{
    Institute of Software, Chinese Academy of Sciences \and
    University of Chinese Academy of Sciences \and 
    Advanced Computing and Intelligence Engineering, Wuxi, China \and
    Beijing Forestry University \\
    \email{gipsyh.icu@gmail.com} \\
    \email{\{qiusong,yiwei\}@iscas.ac.cn} \\
    \email{butianjun24@mails.ucas.ac.cn} \\
    \email{fyy0007@bjfu.edu.cn}
}

\maketitle

\begin{abstract} In this paper, we present rIC3, an efficient bit-level hardware model checker primarily based on the IC3 algorithm. It boasts a highly efficient implementation and integrates several recently proposed optimizations, such as the specifically optimized SAT solver, dynamically adjustment of generalization strategies, and the use of predicates with internal signals, among others. As a first-time participant in the Hardware Model Checking Competition, rIC3 was independently evaluated as the best-performing tool, not only in the bit-level track but also in the word-level bit-vector track through bit-blasting. Our experiments further demonstrate significant advancements in both efficiency and scalability. rIC3 can also serve as a backend for verifying industrial RTL designs using SymbiYosys. Additionally, the source code of rIC3 is highly modular, with the IC3 algorithm module being particularly concise, making it an academic platform that is easy to modify and extend.

\keywords{Formal Verification \and Model Checking \and IC3/PDR.}
\end{abstract}

\section{Introduction}
Model checking \cite{ModelChecking,HandbookMC} is a powerful formal verification technique widely used in modern system design. Given a transition system and a property that describes the desired system behavior, it can efficiently and automatically detect bugs (violations of the property) or prove that the property holds.

IC3 \cite{IC3}, also known as PDR \cite{PDR}, is a prominent SAT-based model checking algorithm widely used in hardware formal verification. It efficiently searches for inductive invariants without unrolling the model. IC3 is distinguished by its completeness in comparison to BMC \cite{BMC}, its scalability compared to IMC \cite{IMC} and K-Induction \cite{KINDUCTION}. Recognized as a state-of-the-art algorithm, IC3 serves as the primary engine for numerous efficient model checkers \cite{ABC,NUXMV,AVR,PONO}.

IC3 has been successfully applied to the verification of industrial designs \cite{IC3IndustrialDesign}. However, it still faces significant challenges due to the state space explosion problem. Therefore, improving the IC3 algorithm and developing more efficient tools to enhance scalability for verifying larger-scale hardware models remains a critical research direction. IC3ref \cite{IC3ref}, the reference implementation of the IC3 algorithm developed by its inventor, serves as a baseline in many IC3-related studies \cite{DeepIC3,Progress,PredictingLemma,IGoodLemma} due to its efficiency and relatively simple codebase. However, it has not been updated in years, and its lack of recent advancements has led to a performance gap. ABC \cite{ABC} provides a highly efficient implementation of the PDR algorithm. However, it is implemented in C and heavily relies on custom data structures, macros, and extensive use of raw pointers, which complicates modification and extension. Similarly, the IC3 implementation in nuXmv \cite{NUXMV} delivers outstanding performance, but its source code is not open-sourced.

In this paper, we introduce rIC3, an bit-level model checker primarily based on the IC3 algorithm. It integrates several optimizations proposed recently. The SAT solver within the IC3 engine is deeply optimized by reducing the number of variables decided through Cone of Influence (COI) analysis \cite{GipSAT}. Furthermore, a more efficient data structure in VSIDS has been implemented to further reduce the solving time for IC3 queries. The tool also leverages CTG \cite{CTG} and EXCTG \cite{DynAMic} to achieve better generalization. Moreover, it employs a heuristic method that dynamically adjusts generalization strategies \cite{DynAMic} to balance the trade-off between generalization quality and computational overhead. Furthermore, the use of internal signals \cite{IC3INN} enables the derivation of more compact invariants.

rIC3 was independently evaluated as the top-performing tool in both the bit-level track and the word-level bit-vector track (via bit-blasting) at the 2024 Hardware Model Checking Competition (HWMCC’24) \cite{HWMCC}. Our experiments further highlight significant advancements in both efficiency and scalability. Additionally, rIC3 can be utilized as a backend for verifying industrial RTL designs with SymbiYosys. The source code of rIC3 is designed with high modularity, and its IC3 algorithm module is remarkably compact, comprising only around 1,700 lines of code. This makes rIC3 an excellent academic platform that is both easy to modify and extend.

\section{Architecture}
\begin{figure}[!t]
    \centering
    \includegraphics[width=0.9\textwidth]{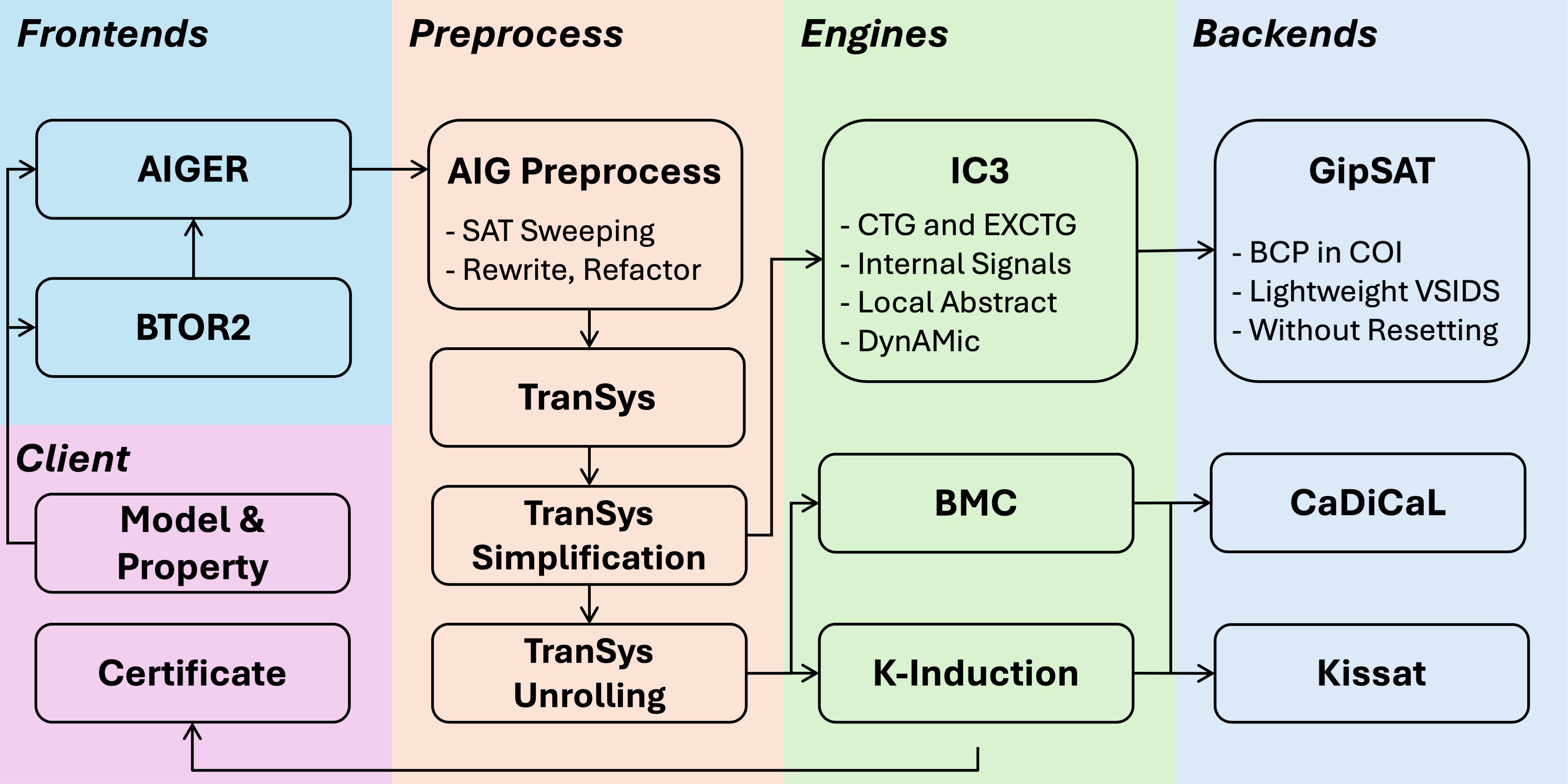}
    \caption{rIC3 Architecture and Verification Flow}
    \label{fig:Flow}
\end{figure}
rIC3 is an efficient bit-level model checker primarily based on the IC3 algorithm, and it also incorporates a portfolio approach with the BMC and K-Induction algorithms. Fig.\ref{fig:Flow} shows the architecture and verification flow of rIC3.

\begin{itemize}
    \item Users provide the model to be verified and the desired properties to the rIC3 frontend, which then obtains the final result certificate from the checker. The certificate is either a proof that the model is safe or a concrete counterexample demonstrating that the model is unsafe.

    \item The frontends in rIC3 support two widely used formats: AIGER \cite{AIGER}, which provides a bit-level representation of hardware, and Btor2 \cite{Btor2}, which uses a word-level bit-vector representation. Since rIC3 operates at the bit level, Btor2 models are bit-blasted into AIGER models using btor2aiger \cite{Btor2AIGER}.

    \item The AIGER model is first preprocessed to obtain a more compact representation, which involves SAT sweeping \cite{SATSweeping}, rewriting \cite{AIGRewrite} and refactoring \cite{AIGFactor}. This step is performed using ABC \cite{ABC}. Subsequently, the model is transformed into TranSys, an intermediate representation of the Transition System, primarily involving CNF encoding. The CNF of TranSys is then further simplified using a SAT solver. Depending on the engine in use, an additional unrolling step may be required.

    \item The engine in rIC3 integrates the IC3, BMC, and K-Induction algorithms, which can also operate in a portfolio mode. These engines formulate SAT problems and delegate them to backend SAT solvers, driving the model checking process based on the results returned. Once a final result is obtained, it is provided to the user along with the corresponding certificate.

    \item The backend supports various SAT solvers, including GipSAT, which excels at solving the relatively simpler SAT problems generated by IC3, as well as CaDiCaL \cite{CaDiCaL} and Kissat \cite{Kissat}, which are more suited for addressing the comparatively complex problems formulated by BMC and K-Induction.

\end{itemize}

\section{Techniques}
Since the BMC and K-Induction are relatively consistent across different model checkers, we focus on introducing the techniques used in the IC3 engine.
\begin{itemize}
    \item \textbf{Specifically optimized SAT solver}. The SAT solver in IC3 primarily focuses on solving the relative induction query $F \land c \land T \land \lnot c'$, where $F$ represents the frame, $c$ is a clause, and $T$ denotes the transition relation. We observed that these queries are relatively simple and are typically solved within milliseconds, exhibiting unique characteristics. Current IC3 implementations mainly rely on modern general-purpose SAT solvers, which offer excellent scalability. However, these solvers may not always be efficient for solving millisecond-level SAT problems and often fail to leverage the unique characteristics of these queries. We developed a novel lightweight SAT solver called GipSAT \cite{GipSAT}, which is specifically optimized for the IC3 algorithm. We observed that the transition relation $T$ has a DAG (Directed Acyclic Graph) structure, meaning it is not necessary to decide and assign all variables during each solving process. GipSAT limits the variables decided and assigned during BCP (Boolean Constraint Propagation) by analyzing the COI (Cone of Influence) at each solving iteration, ensuring that the results remain unaffected. Furthermore, we noticed that the overhead of binary heap operations in the Variable State Independent Decaying Sum (VSIDS) heuristic is not negligible, especially given the simplicity of the queries and the logarithmic time complexity of binary heap operations. To mitigate this, GipSAT uses several buckets to maintain the VSIDS scores, achieving constant-time operations and reducing overhead. Additionally, GipSAT supports temporary clauses, eliminating the need for resetting the SAT solver between queries.

    \item \textbf{Better generalization with CTG and extended CTG}. In IC3, the generalization process aims to expand a bad state to include additional unreachable states, thereby reducing the number of iterations. Standard generalization \cite{Standard} achieves this by removing literals from a cube, effectively encompassing more unreachable states. CTG generalization \cite{CTG} improves upon this approach by blocking counterexamples to generalization (CTG) when dropping literals fails to yield better generalization results. However, the results of CTG may sometimes still be suboptimal. This is because CTG only considers blocking the predecessors of the literal-dropped clause. If blocking these predecessors fails, it abandons further attempts to block the clause, even though the predecessors of the clause's predecessors might still be blockable. To address this limitation, we propose EXCTG \cite{DynAMic}, an extension of CTG. Similar to CTG, EXCTG attempts to block the predecessors when literal dropping fails. However, if blocking the predecessors also fails, EXCTG goes further by recursively attempting to block the predecessors of the predecessors, and so on. This iterative approach enables EXCTG to achieve improved generalization results.

    \item \textbf{Dynamic adjustment of generalization strategies}. While CTG and EXCTG offer better generalization results, they also incur higher computational costs. The generalization strategies: Standard, CTG, and EXCTG, produce progressively better results but also come with increased computational overhead. Finding an appropriate balance between generalization quality and computational overhead is challenging with a static strategy. To address this, we introduce DynAMic (Dynamic Adjustment of MIC Strategies) \cite{DynAMic}, which dynamically adjusts generalization strategies based on the difficulty of blocking bad states, as indicated by the number of failed attempts to block them. The more difficult a bad state is to block, the more effort is devoted to generalization. For states that are easy to block, it uses the lightweight standard strategy to reduce overhead. For more challenging states, it applies more effective generalization strategies, such as CTG or EXCTG, depending on the difficulty. This approach enhances scalability without sacrificing efficiency.

    \item \textbf{Predicate with internal signals}. IC3 incrementally learns inductive invariants that over-approximate the set of reachable states. These invariants are represented in CNF as predicates over state variables. However, IC3 struggles with designs that lack a concise CNF invariant over state variables. To address this limitation, IC3-INN \cite{IC3INN} extends the traditional IC3 approach by learning invariants based not only on state variables but also on internal signals within the design. This enhancement enables the derivation of significantly more compact invariants while maintaining the efficiency of the CNF representation.

    \item \textbf{Localization abstraction with constraints}. Localization abstraction \cite{LocalAbs} is a method aimed at reducing the complexity of a verification instance by removing certain logic. Previous work \cite{IC3LocalAbs} that utilized this method abstracts the transition relation by treating some registers as primary inputs and then refines the abstraction using counterexamples found by IC3. Similarly, rIC3 also takes this approach, but instead of the transition relation, it only focuses on constraints. Initially, it abstracts away all constraints to reduce the complexity. Then, it gradually adds back the constraints that cannot be ignored, based on the counterexamples.
\end{itemize}

\section{Portfolio}
Certain optimizations can significantly improve the solving efficiency of specific model-checking problems. However, achieving consistent performance improvements across all problems remains challenging, as some optimizations may even degrade performance for some other cases. For example, IC3-INN \cite{IC3INN} excels at solving many previously intractable problems, especially SEC problems. However, its overall performance across the entire benchmark set decreases due to treating more variables as latches. Consequently, using a portfolio of configurations in parallel has emerged as a viable strategy to improve scalability. rIC3 includes a 16-thread parallel portfolio combining IC3, BMC, and K-Induction configurations, which is also the competition version. Specifically, it uses 11 threads for IC3 with different combinations of the techniques mentioned above, 4 threads for BMC with varying steps, and 1 thread for K-Induction.

\section{Implementation}
\begin{table}
\centering
\caption{Lines of Code (LOC) for Modules in rIC3}
\label{tab:LOC}
\begin{tabular}{c c c c c c c c c}
\hline
Modules           & IC3   & BMC   & K-Induction & TranSys  & GipSAT & AIG-rs & logic-form & Total \\
\hline  
LOC     &1686   & 115   & 244      & 739          & 2293  & 1654 & 1297 &  10571 \\
\hline
\end{tabular}
\end{table}
All algorithms in rIC3 are implemented in Rust, a modern programming language renowned for its emphasis on performance and security. We have open-sourced rIC3 at \cite{rIC3}. The implementation is designed to be concise while preserving the algorithm’s efficiency, and the codebase is highly modular. Table \ref{tab:LOC} provides statistics on the number of lines of code for various modules in rIC3.

AIG-rs, a self-maintained AIG library, supports AIGER parsing, AIG simulation, and related functionalities. The logic-form library offers data structures for representing basic propositional logic expressions, such as Literal and Cube. Other modules have been introduced earlier. Notably, the IC3 engine demonstrates excellent performance, with its implementation being remarkably simple—comprising only about 1700 lines of code. This simplicity makes it an academic research platform for experimentation and enhancement.

Furthermore, rIC3 is integrated as a backend for SymbiYosys (sby)\footnote[1]{\href{https://github.com/YosysHQ/sby/pull/313}{https://github.com/YosysHQ/sby/pull/313}}, a front-end driver program for Yosys formal hardware verification flows. This integration enables rIC3 to serve as a backend for verifying industrial RTL designs.

\section{Related Work}
IC3 was originally introduced as a SAT-based bit-level model checking algorithm. Since its inception, numerous tools have been developed based on this foundational approach. \textbf{IC3ref} \cite{IC3ref}, created by the algorithm’s inventor, stands out for its efficiency and relatively simple implementation, making it a baseline for many IC3-related studies \cite{DeepIC3,Progress,PredictingLemma,IGoodLemma}. The \textbf{ABC} framework \cite{ABC} includes an implementation of the PDR algorithm \cite{PDR}, while the \textbf{nuXmv} model checker \cite{NUXMV} also integrates IC3. Additionally, \textbf{SimpleCAR} \cite{SimpleCAR} implements the CAR algorithm \cite{CAR}, which excels at bug detection, and \textbf{Avy} \cite{Avy} combines sequence interpolants with IC3.

The IC3 algorithm has also been extended to address word-level problems by leveraging SMT solvers. \textbf{AVR} \cite{AVR} implements IC3sa \cite{IC3SA}, combining IC3 with syntax-guided abstraction to enable scalable word-level model checking. Furthermore, \textbf{Pono} \cite{PONO} supports both IC3sa and IC3ia \cite{IC3IA}, which extends IC3 to modulo theories through implicit predicate abstraction.

The Hardware Model Checking Competition (HWMCC) \cite{HWMCC} has been instrumental in driving advancements in hardware model checkers. Submissions to the competition are typically executed in a 16-thread portfolio mode, prompting many model checkers to develop portfolio versions that integrate various configurations of IC3, BMC, and K-Induction engines. Examples include \textbf{Superprove} \cite{SuperProve} for ABC, \textbf{SuperCAR} \cite{SuperCAR} for SimpleCAR, and \textbf{Pavy} \cite{Pavy} for Avy. AVR and Pono also have their own portfolio versions.

\textbf{rIC3}, like these tools, is used for hardware model checking. It incorporates several optimization techniques proposed in recent years to achieve better scalability, delivering superior performance compared to existing tools. It can also be used to verify industrial models through SymbiYosys. Moreover, it is implemented in the modern programming language Rust, featuring a modular and concise codebase, making it an excellent platform for academic research as well.

\section{Evaluation}
\label{SEC:Evaluation}
\subsection{Setup}
We evaluated the following tools, all of which are the latest versions or the HWMCC'24 submission versions. The specific versions and configurations are detailed in our artifact \cite{Artifact}.
\begin{itemize}
    \item \textbf{rIC3} \cite{rIC3}: rIC3-ic3 is a single-thread IC3 engine that utilizes the DynAMic heuristic method and the specifically optimized SAT solver. Internal signals and localization abstraction are excluded, as they do not enhance overall performance. rIC3-portfolio is the portfolio version of rIC3.
    \item \textbf{IC3ref} \cite{IC3ref}: A single-thread IC3 engine with CTG generalization.
    \item \textbf{ABC} \cite{ABC}: ABC-superprove \cite{SuperProve} is the portfolio engine of ABC. ABC-pdr is the PDR engine using the configuration from ABC-superprove, which includes CTG and localization abstraction.
    \item \textbf{nuXmv} \cite{NUXMV}: We evaluated the IC3 engine in nuXmv, enhanced by a recently proposed heuristic optimization \cite{IGoodLemma}, available at \cite{IGoodLemmaArtifact}. We refer to this as nuXmv-cav23.
    \item \textbf{AVR} \cite{AVR}: AVR-ic3sa is the IC3sa engine using the default configuration, and AVR-portfolio is the portfolio engine of AVR.
    \item \textbf{Pono} \cite{PONO}: Pono-portfolio is the portfolio engine submitted in HWMCC'24. Pono-ic3ia and Pono-ic3sa are engines implementing IC3ia and IC3sa, respectively, using the same configurations as in the HWMCC'24 submission.
    \item \textbf{Avy} \cite{Avy}: Avy is the IC3 engine integrated with interpolants, while Pavy \cite{Pavy} is the portfolio engine that combines Avy, PDR, and BMC.
\end{itemize}

For the evaluation of single-thread engines, we utilized the complete benchmark suites from the bit-level track (AIGER format) and the word-level bit-vector track (Btor2 format) of the three most recent Hardware Model Checking Competitions (HWMCC’19, HWMCC’20, and HWMCC’24) \cite{HWMCC}. The bit-level benchmarks were derived by bit-blasting the word-level bit-vector benchmarks. After removing duplicates, the combined suite comprised 840 unique cases, each available in both AIGER and Btor2 formats. For portfolio engines, due to limited computational resources, we restricted our evaluation to the HWMCC’24 benchmark, which included 319 cases.

All single-thread configurations were tested under consistent resource constraints: 32 GB of RAM and a 3600-second time limit. For the portfolio engine, we followed the HWMCC resource limits, which provided 16 cores, 128 GB of RAM, and the same 3600-second time limit. The experiments were conducted on an AMD EPYC 7532 processor (2.4 GHz) running Ubuntu 24.10. To ensure reproducibility, we have made the source code and binary releases of all evaluated tools available, along with detailed experimental results in \cite{Artifact}. Additionally, to enhance confidence in the correctness of the results, all outputs from rIC3 were certified using Certifaiger \cite{Certifaiger}.

\subsection{Results}

\begin{table}[!t]
\centering
\setlength{\tabcolsep}{9pt}
\caption{Number of solved, timed-out, and memory-out cases, PAR-2 score, number of uniquely solved cases, and fastest solved cases for various IC3 engines. The upper part shows the bit-level engines, and the lower part shows the word-level engines.}
\label{tab:IC3Result}
\begin{tabular}{c c c c c c c}
\hline
Tools & Solved(840) & TO & MO & PAR-2 & Unique & Best \\
\hline
rIC3-ic3 & \textcolor{red}{606} & 225 & 9 & \textcolor{red}{2147.70} & \textcolor{red}{61} & 398\\
nuXmv-cav23 & 533 & 302 & 5 & 2777.30 & 8 & 41 \\
ABC-pdr & 516 & 320 & 4 & 2900.99 & 1 & 80 \\
Avy & 488 & 350 & 2 & 3142.87 & \textcolor{orange}{29} & 38 \\
IC3ref & 486 & 353 & 1 & 3169.29 & 1 & 59 \\
\hline
AVR-ic3sa & 353 & 481 & 6 & 4305.24 & \textcolor{orange}{22} & 53\\
Pono-ic3ia &311 &518 &11 &4652.29 & 1 & 7 \\
Pono-ic3sa & 212 & 614 & 14 & 5459.96 & 0 & 5\\
\hline
\end{tabular}
\end{table}

\begin{figure}[!t]
    \centering
    \includegraphics[width=\textwidth]{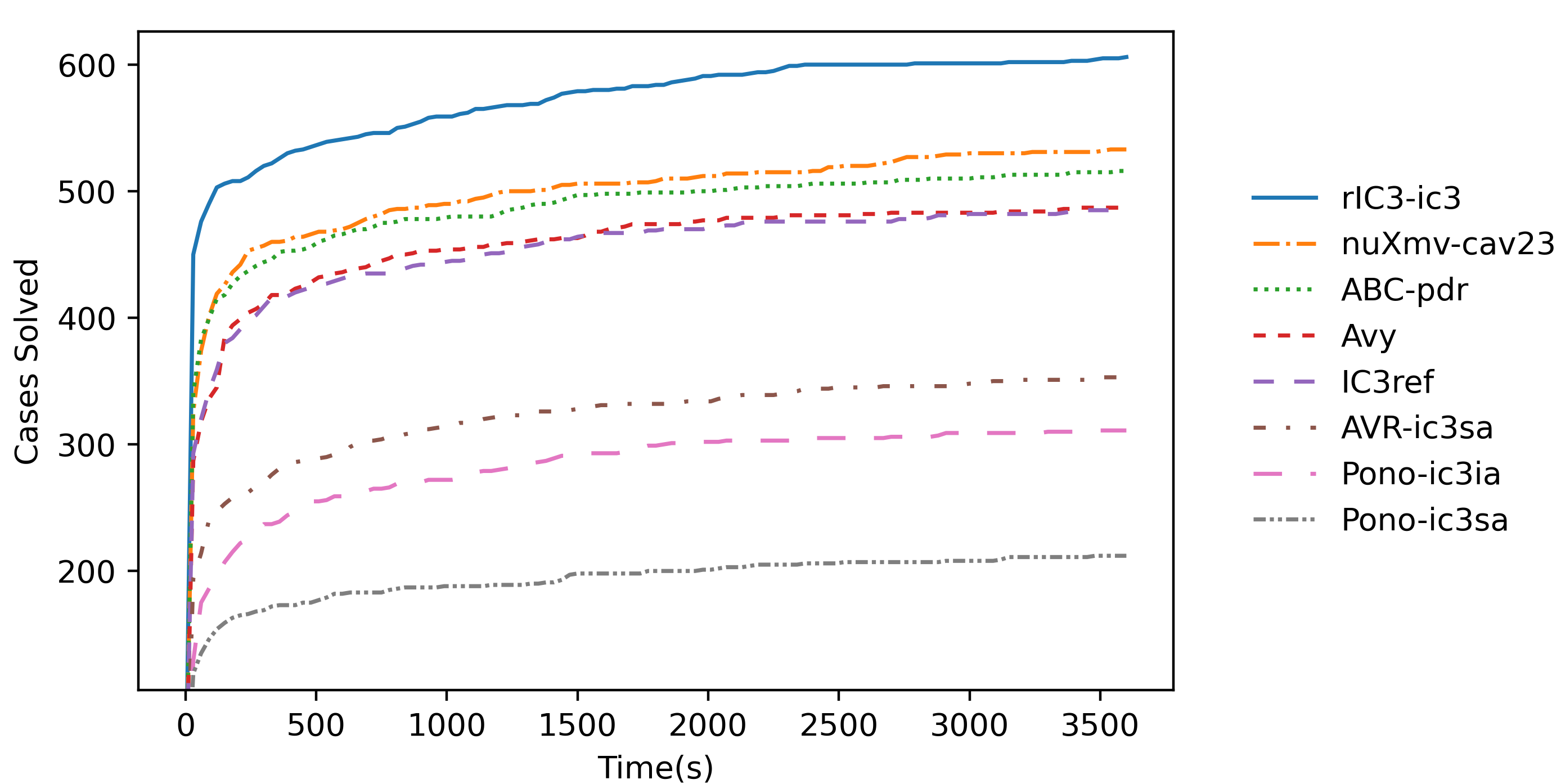}
    \caption{The number of cases solved over time by various single-thread IC3 engines.}
    \label{fig:Plot}
\end{figure}

\begin{figure}[!t]
    \centering
    \includegraphics[width=\textwidth]{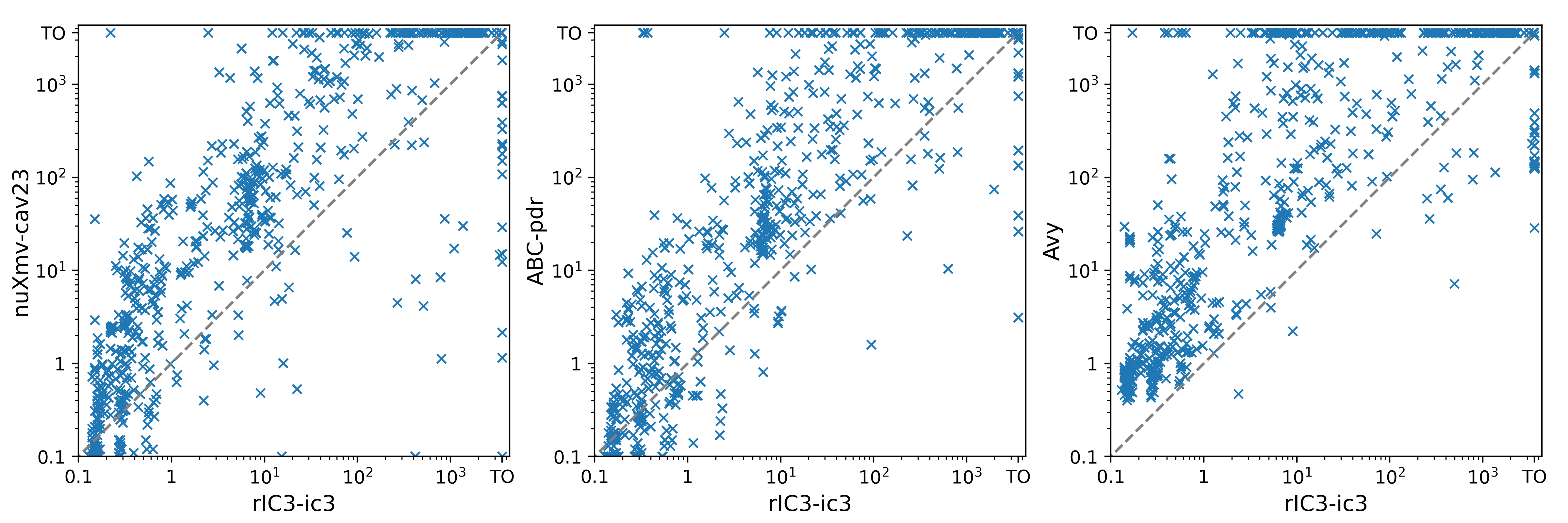}
    \caption{The solving times (in seconds) comparisons between various IC3 engines.}
    \label{fig:Scatter}
\end{figure}

\begin{table}[!t]
\centering
\setlength{\tabcolsep}{9pt}
\caption{The number of solved, timeout, and memory-out cases, the PAR-2 score, the number of uniquely solved cases, and the fastest solved cases for portfolio engines.}
\label{tab:PortfolioResult}
\begin{tabular}{c c c c c c c}
\hline
Tools & Solved(319) & TO & MO & PAR-2 & Unique & Best \\
\hline
rIC3-portfolio & \textcolor{red}{245} & 73 & 1 & \textcolor{red}{1809.74} & \textcolor{red}{16} & 171\\
ABC-superprove & 226 & 92 & 1 & 2197.01 & 2 & 43 \\ 
Pavy\footnotemark[2] & 202 & 116 & 1 & 2801.99 & 0 & 1 \\
AVR-portfolio & 181 & 137 & 1 & 3220.23 & 3 & 22 \\
Pono-portfolio & 156 & 163 & 0 & 3780.85 & 1 & 19\\
\hline
\end{tabular}
\end{table}

\begin{figure}[!t]
    \centering
    \includegraphics[width=\textwidth]{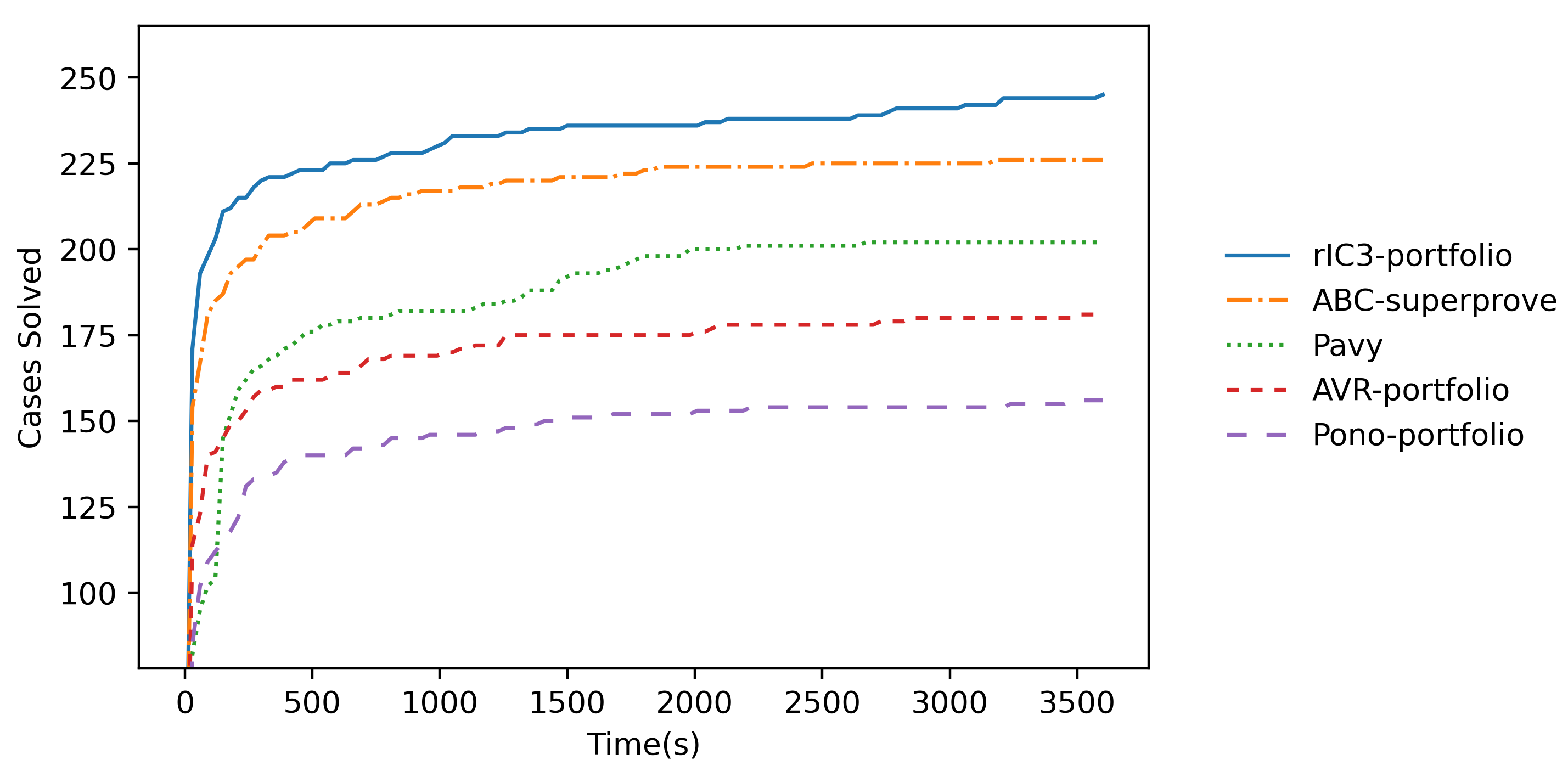}
    \caption{The number of cases solved over time by various portfolio engines.}
    \label{fig:Portfolio-Plot}
\end{figure}

\begin{figure}[!t]
    \centering
    \includegraphics[width=\textwidth]{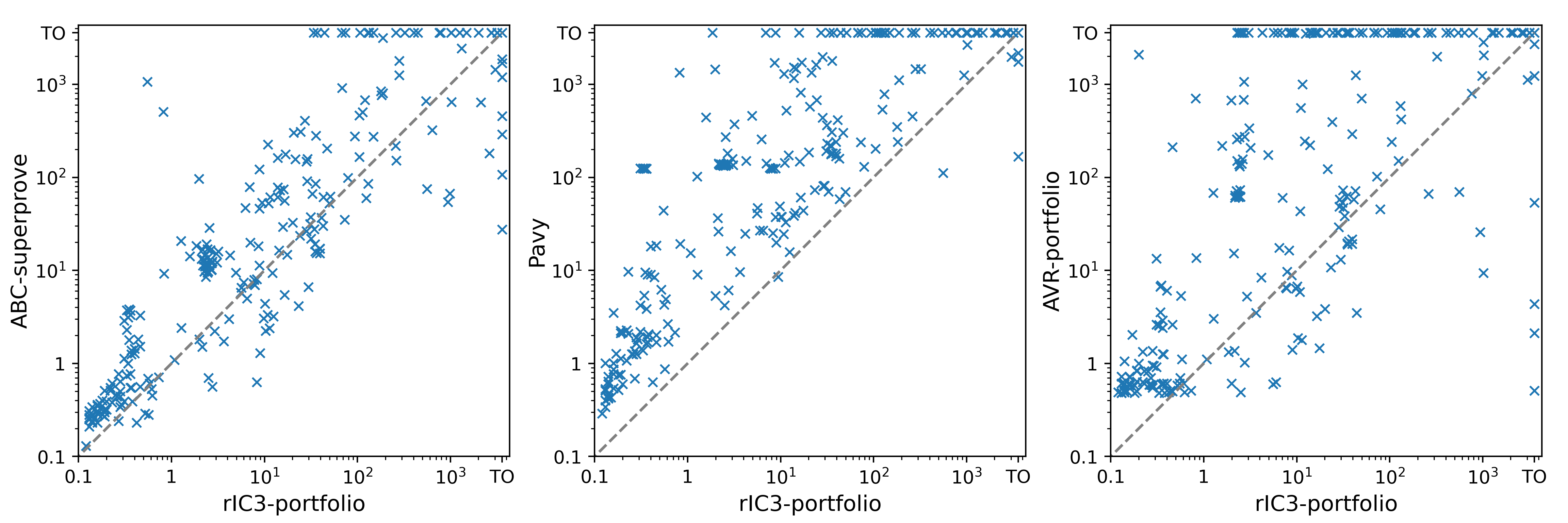}
    \caption{The solving times (in seconds) comparisons between various portfolio engines.}
    \label{fig:Portfolio-Scatter}
\end{figure}

Table \ref{tab:IC3Result}, Figure \ref{fig:Plot}, and Figure \ref{fig:Scatter} present the results of various IC3 engines. The data clearly demonstrate that rIC3 significantly outperforms other engines in terms of scalability. Specifically, rIC3 solves 73 more cases than nuXmv-cav23 and 90 more cases than ABC-pdr. Additionally, rIC3 uniquely solves 61 cases. Both the scatter plot and the PAR-2 score further highlight rIC3’s efficiency, as it solves the vast majority of cases faster than the other engines and achieves the best performance in 398 cases. The results for the portfolio engines are presented in Table \ref{tab:PortfolioResult}, Figure \ref{fig:Portfolio-Plot}, and Figure \ref{fig:Portfolio-Scatter}. It is also evident that the portfolio engine of rIC3 maintains a competitive advantage over the other engines.

\footnotetext[2]{Pavy results differ slightly from HWMCC'24, possibly the version we found \cite{Pavy} differs from the final submitted version.}

To evaluate the effectiveness of the techniques employed, we conducted ablation experiments, as detailed in Table \ref{tab:Ablation}. The default configuration, referred to as rIC3-ic3, incorporates DynAMic and GipSAT but does not utilize internal signals or localization abstraction. To assess the impact of each optimization, we experimented with several variations: replacing GipSAT with Minisat (rIC3-ic3-ms), disabling DynAMic and using CTG (rIC3-ic3-ctg), and enabling internal signals (rIC3-ic3-inn) and localization abstraction (rIC3-ic3-la). The results indicate that disabling either DynaMic or GipSAT led to a reduction in the number of solved cases. Moreover, while enabling internal signals or localization abstraction caused some performance degradation, these configurations uniquely solved certain cases that other setups could not address.

\begin{table}[!t]
\centering
\setlength{\tabcolsep}{9pt}
\caption{The number of solved, timeout, and memory-out cases, the PAR-2 score, and the number of uniquely solved cases for various configurations.}
\label{tab:Ablation}
\begin{tabular}{c c c c c c c}
\hline
Tools & Solved(840) & TO & MO & PAR-2 & Unique \\
\hline
rIC3-ic3 & 606 & 225 & 9 & 2147.70 & 13 \\
\hdashline
rIC3-ic3-ms & 564 & 275 & 1 & 2492.34 & 1 \\
rIC3-ic3-ctg & 590 & 242 & 8 & 2254.30 & 6 \\
rIC3-ic3-inn & 564 & 274 & 2 & 2447.95 & 12 \\ 
rIC3-ic3-la & 605 & 226 & 9 & 2173.03 & 25 \\
\hline
\end{tabular}
\end{table}

\section{Strengths and Limitations}
As a bit-level model checker, rIC3 demonstrates significant performance advancements, outperforming not only other bit-level checkers but also those leveraging word-level information. Its implementation is highly modular, with the IC3 algorithm module being particularly concise. However, due to its lack of utilization of word-level information, rIC3 is relatively inefficient in handling cases involving arithmetic logic. Moreover, it currently cannot solve bit-vector problems involving arrays, as bit-blasting an array model can lead to an explosion. Additionally, as shown in Table \ref{tab:IC3Result}, AVR and Avy uniquely solved many cases that rIC3 could not, as their algorithms differ substantially from the original IC3 algorithm. Therefore, incorporating more algorithm variants into rIC3's portfolio engine could further enhance its performance.

\section{Conclusion}
In this paper, we introduce rIC3, a novel hardware model checker. Both the HWMCC’24 results and our experimental evaluations highlight its significant advancements over existing tools. rIC3 can also function as a backend for verifying industrial RTL designs using SymbiYosys. Furthermore, the source code of its IC3 algorithm module is highly concise, making rIC3 an ideal academic platform that is easy to modify and extend. Looking ahead, we plan to enhance rIC3 by incorporating word-level information to further boost its performance and to enable support for solving word-level problems involving arrays.

\section*{Acknowledgement}
This work was supported by the Beijing Municipal Natural Science Foundation (Grant No. 4252024), the Foundation of Laboratory for Advanced Computing and Intelligence Engineering (Grant No. 2023-LYJJ-01-013), the Basic Research Projects from the Institute of Software, Chinese Academy of Sciences (Grant No. ISCAS-JCZD-202307) and the National Natural Science Foundation of China (Grant No. 62372438).

\subsubsection*{Disclosure of Interests}
The authors have no competing interests to declare that are relevant to the content of this article.

\bibliographystyle{splncs04}
\bibliography{bib}
\end{document}